# Distributed Periodic Approach for Adaptive Fault Diagnosis in Distributed Systems

**Latika Sarna**
(Cummins College of Engineering for Women, Pune, India
latika.sarna@cumminscollege.in)

**Sumedha Shenolikar**
(Cummins College of Engineering for Women, Pune, India
sumedhashenolikar28@gmail.com)

**Poorva Kulkarni**
(Cummins College of Engineering for Women, Pune, India
poorva.kulz@gmail.com)

**Varsha Deshpande**
(Cummins College of Engineering for Women, Pune, India
vdvdeshpande@gmail.com)

**Supriya Kelkar**
(Cummins College of Engineering for Women, Pune, India
supriya.kelkar@cumminscollege.in)

**Abstract:** In this paper, Decentralized Periodic Approach for Adaptive Fault Diagnosis (DP-AFD) algorithm is proposed for fault diagnosis in distributed systems with arbitrary topology. Faulty nodes may be either unresponsive, may have either software or hardware faults. The proposed algorithm detects the faulty nodes situated in geographically distributed locations. This algorithm does not depend on a single node or leader to detect the faults in the system. However, it empowers more than one node to detect the fault-free and faulty nodes in the system. Thus, at the end of each test cycle, every fault-free node acts as a leader to diagnose faults in the system. This feature of the algorithm makes it applicable to any arbitrary network. After every test cycle of the algorithm, all the nodes have knowledge about faulty nodes and each node is tested only once. With this knowledge, there can be redistribution of load, which was earlier assigned to the faulty nodes. Also, the algorithm permits repaired node re-entry and new node entry. In a system of *n* nodes, the maximum number of faulty nodes can be (*n*-1) which is detected by DP-AFD algorithm. DP-AFD is periodic in nature which executes test cycles after regular intervals to detect the faulty nodes in the given distributed system.

**Keywords:** distributed; fault diagnosis; fault detection; computer networks; distributed algorithm; adaptive; periodic; arbitrary networks

**Categories:** C.2.4, C.2.5, C.2.6

# 1 Introduction

It becomes extremely important to detect the availability of the computers in distributed systems. There are different proposed systems to detect and handle faulty nodes. The software faults may arise due to faults such as memory failures, arithmetic or logical faults found in the processors. Also, hardware fault may be present due to communication link or input-output hardware interface failures. The diagnosis algorithm for the faulty nodes should be capable of discovering all the faulty nodes in the system and should not diagnose any fault-free node as faulty. The algorithm is adaptive if it can adapt to different fault conditions along with different topological schemes and accordingly be able to modify the testing strategies. There are different models proposed with either a single central observer or a coordinator pair to detect faults. Also, there are certain peer to peer model based fault detection schemes which may suffer from communication overhead. Decentralized Periodic Approach for Adaptive Fault Diagnosis does not have a one-point or two-point of failures and works in peer-to-peer fashion using minimal number of messages thereby reducing the communication overhead.

# 2 Related Work

Work done by [Ziwich, Duarte (16)] proposes a comparison based model to give a nearly optimal algorithm. This algorithm reduces the time complexity of previously presented algorithms based on arbitrary topology. Its diagnosis is based on comparing the outputs returned by a system pair. This work introduces a new comparison based diagnosis to identify faults in a t-diagnosable system of arbitrary topology under the MM* model.

Coordinator based adaptive fault diagnosis algorithm presented by [Kelkar, Yeole, Sinkar, Jagtap, Zagade (17)] for distributed networks proposes a technique where instead of a single coordinator supervising the entire system, a coordinator pair is used for finding faulty nodes. This algorithm is periodically executed for every node. It also allows the repaired nodes to re-enter the system. ($n$-2) number of nodes can be detected as faulty nodes by this algorithm.

[Kelkar, Kamal (14)] have proposed an algorithm for detecting faulty nodes on Controller Area Network. This algorithm proves that it utilizes a specific number of test rounds and messages to diagnose faults in the system. However, this algorithm is deployed only for static and automotive networks. New nodes and repaired faulty nodes can enter into the system during the diagnostic cycle of the algorithm.

The Leader based fault diagnosis algorithm proposed by [Manghwani, Taware, Kelkar, Chinde, Alwani (17)] works by selecting a leader. The other nodes send their diagnostic information to the leader node. Based on whether the packet is received or not, the node is classified as faulty or faulty-free and this information is maintained by the leader.

[Zhao, Liu, Liu, He, Wang, Wang (17)] have proposed a new technique for fault diagnosis involving a graph based approach. Here, faults are diagnosed by applying clustering on similar test failures. This technique uses a multi-relational graph based approach which is better than outmoded methods for both simulation and monitoring.

[Marino, Pierri (15)] have proposed an algorithm based on discrete linear time aspects. The algorithm solves the issue of non-federated control and diagnosis for groups of cooperative autonomous vehicles.

The algorithm proposed by [Zhu, Jue, Ying, Jianbo (17)] implements an efficient fault diagnosis technique using Zigbee protocol. In order to curb the costly hardware required to detect faults, and to overcome inadequacies of time-honored fault diagnosis systems, particle filter technology is used.

The work proposed by [Rahme, Meskin (17)] presents a novel technique in which a distributed adaptive sliding mode observer scheme is developed. This technique mainly focuses on fault diagnosis in large-scale non-linear dynamical networks. The observer's synthesis relies on each individual nodes dynamics.

[Keroglou, Hadjicostis (18)] have proposed an approach to verify diagnosability for a class of set intersection refinement strategies. This can be used in fault diagnosis of non-deterministic finite automata. Strategies used here help to communicate the diagnostic information with other observational sites periodically. Depending on the diagnostic information such as normal, operational or faulty, the operational sites continue the operation based on the refined diagnostic information.

Failure diagnosis in distributed systems using targeted fault injection has been proposed by [Pham, Wang, Tak, Baset, Tang, Kalbarczyk (16)]. This technique uses a database based failure diagnosis approach. This database is populated by injecting the faults in the construction phase. The faults are queried and matched with the most similar fault in the database.

[Qi, Yao, Uzunov (17)] have proposed Fault Detection and Localization in Distributed Systems Using Recurrent Convolutional Neural Networks. This system uses an automated fault detection methodology having a database with CPU and network usage which is trained using a data interpreter and RCNN-based model. The fault classifier then classifies the types of faults based on the training.

Chen and Cong have proposed a fault detection and isolation algorithm via deterministic learning [Hu, Wang, Dong (14)]. It learns about the connections and fault functions. Using this knowledge some estimates are constructed for each subsystem and then the analysis of its detection and isolation capabilities are made.

[Duarte, Bona, Ruoso (14)] in their work have proposed another distributed diagnosis approach, called VCube. It is an algorithm for virtually interconnecting network nodes. Considering no faults in the system, the fundamental principle of VCube is that it becomes a logical hypercube having many logarithmic properties.

[Ferdowsi, Jagannathan, Zawodniok (14)] have presented a new approach to fault detection using Outlier analysis. Outlier analysis is the process of identifying anomalies or inconsistencies in the data which can be a contemporary solution to fault diagnosis. In order to evaluate the actual system state and simultaneously to identify and remove the faults, a feed forward neural networks (NN) has been considered.

A distinct distributed fault diagnosis is suggested by [Duarte, Weber, Fonseca (12)]. In this paper, Distributed Network Reachability (DNR) algorithm has been proposed which demonstrates the nodes which are reachable and unreachable. The topology is arbitrary and dynamic, and the faults considered are crash faults as well as timing faults.

[Punyotoya, Khilar (10)] have proposed a fault diagnosis algorithm for distributed clusters. The network is considered to have an arbitrary topology with *k* connectivity, where *k* designates the number of clusters. There are series of intermediate nodes which send messages between the desired source and destination.

As described above, there are various approaches for fault diagnosis. Few approaches focus on fault diagnosis in arbitrary networks mainly [Ziwich, Duarte (16)], [Duarte, Weber, Fonseca (12)] and [Punyotoya, Khilar (10)], few other approaches as proposed in [Kelkar, Yeole, Sinkar, Jagtap, Zagade (17)], [Manghwani, Taware, Kelkar, Chinde, Alwani (17)] and [Rahme, Meskin (17)] highlight fault diagnosis using either a leader or an observer. Some other focus on using database-based approach for identifying faults mainly [Pham, Wang, Tak, Baset, Tang, Kalbarczyk (16)] and [Qi, Yao, Uzunov (17)], some mentioned in [Hu, Wang, Dong (14)] and [Ferdowsi, Jagannathan, Zawodniok (14)] use artificial intelligence for detecting faults in a system. The methods presented in [Kelkar, Kamal (14)] and [Marino, Pierri (15)] focus on fault diagnosis in autonomous vehicles and automotive networks, while fault diagnosis can also be performed involving graph based approach as given in [Zhao, Liu, Liu, He, Wang, Wang (17)] and also using Zigbee protocol as given in [Zhu, Jue, Ying, Jianbo (17)]. Some other studies also show that fault diagnosis can carried out using Finite Automata presented in [Keroglou, Hadjicostis (18)]. Another fault diagnosis algorithm is proposed in [Duarte, Bona, Ruoso (14)] for virtually interconnected networks called VCube.

Considering the various approaches presented, DP-AFD tries to carve a niche of its own by adapting dynamically to any arbitrary network without restrictions on the topology. In DP-AFD, each node is tested only once, which helps in preventing redundant node checking, which in turn reduces the number of messages. This makes it efficient. The parallel dissemination of results to all nodes makes the algorithm faster.

## 3  DP-AFD

### 3.1  System Details and Assumptions

In DP-AFD, no node has prior knowledge about the network. However, each node must know its neighbours. The proposed system is assumed to be an arbitrary network, as shown in [see Fig. 1]. DP-AFD assumes the system to be not fully-connected i.e. every node is not connected to every other node. Therefore, some nodes need to interact with intermediate nodes for establishing communication with each other. Examples of arbitrary networks include internet and intranets. Let G be the set containing the nodes in a system, where $N_1, N_2, .., N_n$ are the nodes in the system then,

$$G = \{N_1, N_2, .., N_n\} \qquad (1)$$

Where,

$n$ = Total number of nodes in the system.

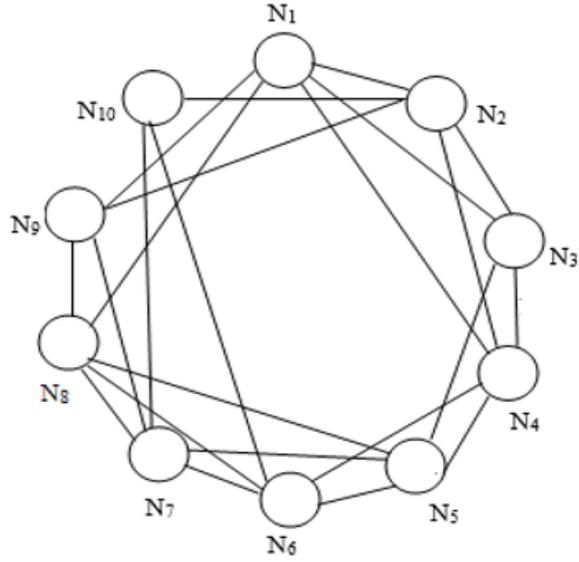

*Figure 1: Proposed System with arbitrary topology used for fault diagnosis*

Assumptions:
1) Algorithm is applicable to arbitrary network.
2) The network is not fully connected.
3) Number of nodes in the network is not known. A node is aware of only its neighbours.
4) The node is assumed to be fault-free, if the computed result is correct.
5) Communication link failures and node failures are not differentiated.
6) A faulty node can participate in the next diagnosis cycle after getting repaired in the current diagnosis cycle.
7) Faults can be temporary or permanent.

[see Fig. 1] shows the proposed system used for implementing the DP-AFD algorithm for fault diagnosis. DP-AFD uses the following fault model for the arbitrary network. In DP-AFD, every node $n_i$ performs a self-test initially. Node $n_i$ sends its self-test result, $r$, through acknowledgement frame [see Fig. 2] to neighbour node $n_j$. The received self-test result is stored in buffer of node $n_j$ as $B$. The fault model determines if the node is faulty or fault-free by comparing the self-test result sent through the acknowledgement frame.

$$f(r, B) = \begin{cases} 0, & \text{fault-free} \\ 1, & \text{otherwise} \end{cases}$$

Where $r$ is result of self-test performed by $n_i$ and $B$ is result stored in a buffer of $n_j$.

$f(r, B)$ equal to 0 denotes $n_i$ as fault-free and 1 denotes $n_i$ as faulty. If acknowledgment frame is not received by $n_j$, then $n_i$ is considered as faulty node.

| IP Address | Status | Leader Bit |
|---|---|---|

*Figure 2: Local Frame and Acknowledgement Frame*

### 3.2  Details of DP-AFD Algorithm

The DP-AFD algorithm consists of two main stages.
1. Leader election stage
2. Fault diagnosis stage.

The 'leader election stage' is executed in four phases namely
 a) Self-test, check and broadcast phase
 b) Acknowledgement phase
 c) Comparison phase
 d) Election phase.

The second stage, the 'fault diagnosis stage' consists of four phases mainly
a) Request and response phase
b) Updation of result frame phase
c) Next leader selection phase
d) Reporting of faulty nodes phase.

### 3.2.1  Leader Election Stage

In order to begin the fault diagnosis cycle, it is necessary to have a node which acts as an initiator of the execution. This node which acts as an initiator is elected using the leader election stage. Following is the detailed explanation of each phase of the leader election stage.

*3.2.1.1 Self- test, check and broadcast phase*:

Initially, every node starts its self-test cycle. Self-test consists of several tests such as input-output, floating point, arithmetic, and memory operations. After the completion of self-test, every node appends its fault status locally in the form of local frame as shown in [see Fig. 2]. If the node finds itself as fault-free, it checks if it has received broadcast message namely, $B_{Cast}$, from any other node [see Fig. 3]. This broadcast message sent by any other node is about that node volunteering to be the leader.

a) If any node receives a broadcast message after the self-test, it will not volunteer for leadership but simply acknowledges the broadcast message using acknowledgement frame. This is depicted as ACK in [see Fig. 3].
b) If the node after the self-test does not receive any broadcast, $B_{Cast}$, within a predetermined time, then it will broadcast the message indicating to be the volunteer for becoming the leader.

*3.2.1.2 Acknowledgement Phase*:

All the nodes which receive the broadcast from the node volunteering to become the leader, send the acknowledgement to it along with its status using the acknowledgement frame. This acknowledgement frame is nothing but a copy of local frame [see Fig. 2], which is sent by the neighbour node to the volunteering node. The acknowledgement frame contains three fields, namely, the IP address of the node sending the acknowledgement, its fault status as 0 or 1 and whether the node has already become the leader. Initially, the leader bit will be 0. As earlier mentioned, every node knows its neighbours. When any node volunteers to become the leader, it sends a broadcast message to its neighbours. If any of its neighbours do not respond in the given prescribed time, the volunteering node considers that neighbour node as faulty and updates its status as 1 indicating that particular neighbour node to be faulty.

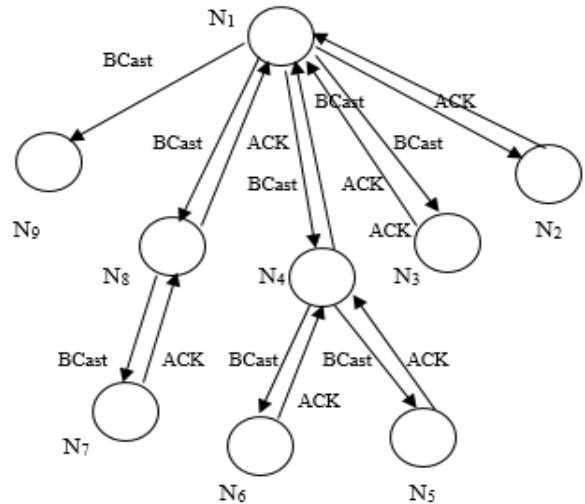

*Figure 3: Request and Acknowledgement Messages*

### 3.2.1.3 Comparison Phase:

There may be a situation where, a node receives broadcast from more than one node volunteering to become the leader. In this case, the node receiving the broadcast messages will acknowledge to the volunteering node whose broadcast message it has received first.

### 3.2.1.4 Election Phase:

At the end of the leader election stage, there will be only one leader, even if there is more than one node volunteering to be the leader. The election of the leader is made on the basis of maximum number of acknowledgements received by each of the volunteering nodes. If two or more volunteering nodes receive equal number of acknowledgements then the volunteering node which broadcasts its status at the earliest as leader will be elected as the leader. Therefore, in order to accomplish this, there will be communication between all the volunteering nodes [see Fig. 4]. This communication could be direct or via the other nodes connected between them. These volunteering nodes send number of acknowledgements they have received to each other. Now, every volunteering node compares the number of acknowledgements received by itself with the acknowledgement received by other volunteers. The node having the maximum number of acknowledgements will become the leader. It broadcasts its status as the leader to all the nodes of the system using the local frame by setting the leader bit as 1 as shown in [see Fig. 2].

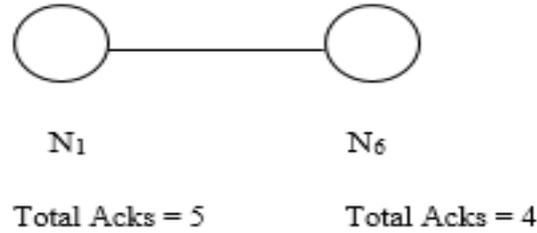

Figure 4: Acknowledgement Comparison

### 3.2.2 Fault Diagnosis Stage
The fault diagnosis stage consists of 4 phases namely
1. Request and response phase
2. Updating result frame at leader node
3. Next leader selection
4. Reporting faulty nodes
Following is the detailed explanation of each phase in the fault diagnosis stage of DP-AFD.

*3.2.2.1 Phase 1: Request and response phase:*
    *3.2.2.1.1* Elected leader will send the request frame which is a copy of the local frame [see Fig. 2] to each of its neighbours. This request frame contains leader bit set as 1.
    *3.2.2.1.2* Each neighbour will send its fault status with the help of acknowledgement frame. This fault status is based on the self-test. The above said acknowledgement frame sent by the neighbour has the leader bit set as 0. This is because this neighbour node has not yet been explored and hence not yet become the leader. As shown in [see Fig. 6(a)], at the end of this phase, leader node $N_1$ will have the fault status of all its neighbours namely $N_2$, $N_3$, $N_4$ and $N_8$ and $N_9$.

| Leader Bit | Leader$_i$ | Leader$_j$ | ... | ... | Leader$_n$ |
|---|---|---|---|---|---|
| IP Address | IP$_i$ | IP$_j$ | ... | ... | IP$_n$ |
| Status Bit | Status$_i$ | Status$_j$ | ... | ... | Status$_n$ |
| Node | N$_i$ | N$_j$ | ... | ... | N$_n$ |

Figure 5: Result Frame

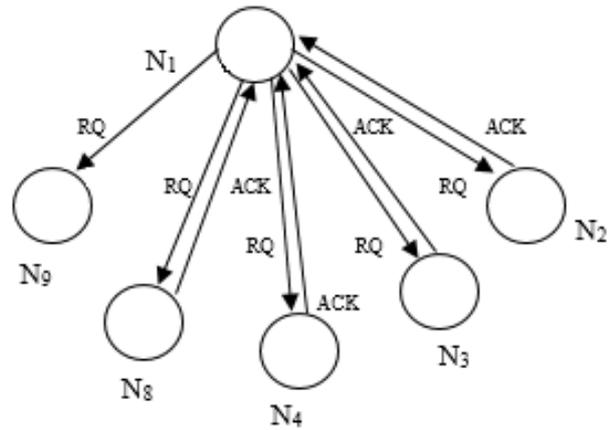

(a)

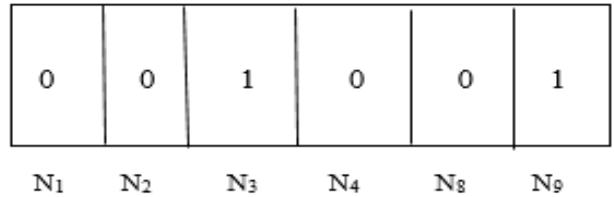

(b)

*Figure 6: Request and Acknowledgement with Result Frame with Status bit*

*3.2.2.2  Phase 2: Updating Result Frame at leader node*

The leader node extracts the fault status sent in the form of acknowledgement by the neighbour node. The result frame [see Fig. 5] is appended by the leader after the receiving the acknowledgement from each neighbour. [see Fig. 6(a)] shows the communication between the leader and its neighbour. [see Fig. 6(b)] shows the status bits of the neighbour nodes stored in the result frame at the leader $N_1$. Status bit 0 indicates the respective neighbour node is fault-free and status bit 1 indicates the respective neighbour node as faulty. Thus, as shown in [see Fig. 6(b)], node $N_3$ and $N_9$ are found to be faulty. This is due to no response received from $N_9$ in the given prescribed time and self-test failure received from $N_3$ by the leader $N_1$.

*3.2.2.3  Phase 3: Next Leader Selection*

*3.2.2.3.1* In DP-AFD algorithm, the current leader selects the next leader. The fault-free node with the least value of response time is selected as the next leader only if the leader bit in result frame of that node is found to be 0 by the current leader. Now the current leader sends its updated result frame to the next chosen leader.

*3.2.2.3.2* The new selected leader will repeat the entire fault diagnosis stage for the new and unexplored neighbours, which were not part of the earlier fault diagnosis stage. This ensures that each node is tested only once by the leader.

*3.2.2.3.3* If there are no new neighbours discovered by the new selected leader, then algorithm ensures that previous leader is given the control. The previous leader now selects another fault-free node with leader bit equal to 0 as a new leader. This new leader is the one, whose response time is next best of what was found for earlier new selected leader.

*3.2.2.3.4* This process continues till there are no unexplored nodes left in the system. Thus, all the fault-free nodes are given a chance to become the leader. This scheme allows the network to be discovered dynamically.

*3.2.2.4 Phase 4: Reporting Faulty Nodes*

As discussed earlier, the most recent leader has the fault status information about all the nodes in network. This is stored in the form of result frame at the recent leader. Therefore, now the recent leader will extract the fault status and the IP of only the faulty nodes from the result frame and broadcasts the information to all the nodes in the network. As shown in [see Fig. 7], $N_5$ broadcasts the faulty status of nodes namely, $N_3$, $N_6$ and $N_9$ to all the nodes in the network. This is the end of one diagnostic cycle in DP-AFD algorithm.

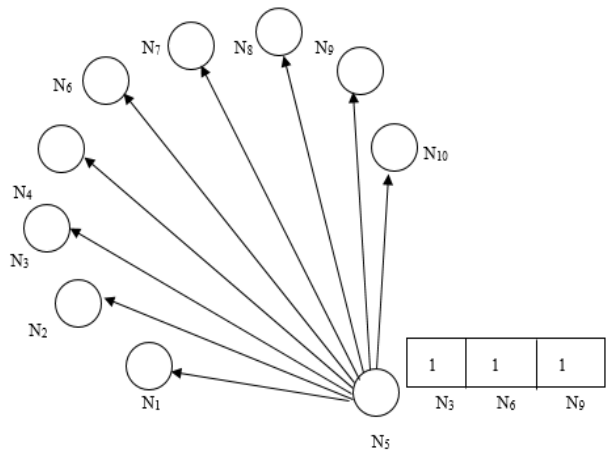

*Figure 7: Final Broadcast*

DP-AFD algorithm is executed periodically. DP-AFD dynamically detects the network while diagnosing the faults. It is also proved that the *t*-diagnosability of DP-AFD is ($n$-1). This means that DP-AFD successfully works even if there is only one fault-free node in the network.

**3.3 New Node Entry and Repaired Node Re-entry**

Whenever a new node enters or repaired node re-enters the network, it is expected to perform the self-test and update its local frame. New node or repaired node is diagnosed by DP-AFD in the next fault diagnosis cycle.

## 4 Analysis of DP-AFD

Calculation of total number of messages in one diagnosis cycle can be given as follows:

a) In the request-response phase, request message will be sent to all the neighbour nodes of the current leader.

Number of messages sent in this phase namely $M_R$

$$M_R = \deg(l) \tag{2}$$

Where,
$\deg(l)$ = degree of the current leader or number of nodes connected to it

b) In the request response phase, acknowledgement messages will be received by the current leader from the neighbour nodes namely $M_A$

$$M_A = \deg(l) - f_n \tag{3}$$

Where,
$f_n$ = number of faulty neighbour nodes

c) In the next leader selection phase, the result frame is sent by the current leader to the next new selected leader namely $M_{RE}$

$$M_{RE} = 1 \tag{4}$$

d) At the end of diagnostic cycle of DP-AFD, the recent leader broadcasts the information of the faulty nodes to all the nodes of the system. This broadcast message namely $M_{BCast}$ is,

$$M_{BCast} = n - 1 \tag{5}$$

Where,
$n$ = number of nodes

e) Sum total of all the above messages in a, b, c, d is:
$M_{total} = M_R + M_A + M_{RE} + M_{BCast}$
Therefore,
$$M_{total} = 2 * \deg(l) - f_n + n \tag{6}$$

Where,

$M_{total} =$ total number of messages in DP-AFD algorithm for a single leader,

f) Therefore when there are two or more leaders in the given system, the total number of messages are:

$$M_{cycle} = \left[ \sum_{i=1}^{N(L)} 2\deg(l_i) - f_n \right] + n \qquad (7)$$

Where,
$N(L) =$ number of leaders chosen in the entire network
$\deg(l) =$ degree of the current leader or number of nodes connected to it
$f_n =$ number of faulty neighbour nodes
$n =$ number of nodes in the system

## 5 Results

DP-AFD is implemented on two VLANs where each VLAN with ten computer systems. The two VLANs interact with each other to exchange information about the faulty nodes in their respective networks. The [see Tab. 1] shows the IP addresses of the nodes in each VLAN.

| Nodes | VLAN 20 | VLAN 30 |
|---|---|---|
| $N_1$ | 172.16.20.101 | 172.16.30.101 |
| $N_2$ | 172.16.20.102 | 172.16.30.102 |
| $N_3$ | 172.16.20.103 | 172.16.30.103 |
| $N_4$ | 172.16.20.104 | 172.16.30.104 |
| $N_5$ | 172.16.20.105 | 172.16.30.105 |
| $N_6$ | 172.16.20.106 | 172.16.30.106 |
| $N_7$ | 172.16.20.107 | 172.16.30.107 |
| $N_8$ | 172.16.20.108 | 172.16.30.108 |
| $N_9$ | 172.16.20.109 | 172.16.30.109 |
| $N_{10}$ | 172.16.20.110 | 172.16.30.110 |

*Table 1: IP address and Node mapping of 2 VLANs nodes*

*(a) Volunteers for becoming the first leader*

*(b) First leader selected*

Figure 8: More than one leader volunteers

[see Fig. 8] and [see Fig. 9] indicate different stages in a fault diagnosis cycle of DP-AFD.

**A. Scenario 1: Two nodes volunteering to become leaders**

[see Fig. 8] shows election phase of leader election stage. Here node $N_8$ and node $N_4$ are volunteering to become leaders with IP addresses 172.16.20.108 and 172.16.20.104 respectively in VLAN 20 [see Fig. 8(a)]. One of them will be leader based on maximum number of acknowledgements received from the neighbour nodes. If the acknowledgements received by $N_8$ and $N_4$ are found to be same, the election of the leader will be done as per election phase discussed earlier in [see Section 3.2.1.4]. Here, $N_8$ is receiving more acknowledgements compared to $N_4$. Hence, it becomes the first leader as shown in [see Fig. 8(b)].

*Figure 9: Next Leader Elected*

### B. Scenario 2: Selection of next leader

[see Fig. 9] shows the next leader election phase under fault diagnosis stage for the VLAN 20.This screenshot is captured at node $N_5$ with IP address 172.16.20.105. The data_lst shown in [see Fig. 9] indicates the previous leader $N_7$ with IP address 172.16.20.107, sending its result frame to the new elected leader, $N_5$ with IP address 172.16.20.105. Data_lst shows $N_7$ as the leader and $N_5$ receiving the result frame. This indicates that $N_5$ subsequently becomes the new selected leader. This selection is based on the quickest response time given by $N_5$ to $N_7$ in request response phase of fault diagnosis stage.

*Figure 10: Intra VLAN faulty nodes and Repaired Node Re-entry*

**C. Scenario 3: Detecting faulty nodes in VLAN 30 and reentry of repaired node**

The screenshot in [see Fig. 10] shows two diagnosis cycles of the DP-AFD from the log file captured in VLAN 30. In the first diagnosis cycle, $N_9$ with IP address 172.16.30.109 is faulty as its status is found to be 1. In the next diagnosis cycle, $N_9$ is found to be repaired. Therefore it is not part of the faulty list containing faulty nodes in cycle number 2.

```
RESULT FRAME in BACKTRACK RECEIVE FOR WRITING IN PICKLE FORM

[{'status': 0, 'ip': u'172.16.20.109', 'leader': 1}, {u'status': 0, u'ip': u'172.16.20.108', u'leader': 1}, {u'status': 0, u'ip': u'172.16.20
7', u'leader': 1}, {'status': 1, 'ip': '172.16.20.101', 'leader': 0}, {'status': 1, 'ip': '172.16.20.102', 'leader': 0}, {u'status': 0, u'ip'
'172.16.20.106', u'leader': 1}, {u'status': 0, u'ip': u'172.16.20.105', u'leader': 1}, {u'status': 0, u'ip': u'172.16.20.104', u'leader': 1},
status': 1, 'ip': '172.16.20.110', 'leader': 0}, {'status': 1, 'ip': '172.16.20.103', 'leader': 0}]
contents ['172.16.20.105', '172.16.20.107', '172.16.20.104', '172.16.20.108', '172.16.20.110', '']
```
(a)

```
Entry:   172.16.20.109 was already a leader so cant make it a leader.
Entry:   172.16.20.108 was already a leader so cant make it a leader.
Entry:   172.16.20.107 was already a leader so cant make it a leader.
Entry:   172.16.20.101 was faulty so cant make it a leader.
Entry:   172.16.20.102 was faulty so cant make it a leader.
Entry:   172.16.20.106 was already a leader so cant make it a leader.
Entry:   172.16.20.105 was already a leader so cant make it a leader.
Entry:   172.16.20.104 was already a leader so cant make it a leader.
Entry:   172.16.20.110 was faulty so cant make it a leader.
Entry:   172.16.20.103 was faulty so cant make it a leader.
all_leaders_done 10
```
(b)

```
ALL NODES EXPLORED.BROADCAST FAULTY NODES OF RESULT FRAME: [{'status': 0, 'ip': u'172.16.20.109', 'leader': 1}, {u'status': 0, u'ip': u'172.16
0.108', u'leader': 1}, {u'status': 0, u'ip': u'172.16.20.107', u'leader': 1}, {'status': 1, 'ip': '172.16.20.101', 'leader': 0}, {'status': 1,
ip': '172.16.20.102', 'leader': 0}, {u'status': 0, u'ip': u'172.16.20.106', u'leader': 1}, {u'status': 0, u'ip': u'172.16.20.105', u'leader':
, {u'status': 0, u'ip': u'172.16.20.104', u'leader': 1}, {'status': 1, 'ip': '172.16.20.110', 'leader': 0}, {'status': 1, 'ip': '172.16.20.103
 'leader': 0}]
Faulty nodes are [{'status': 1, 'ip': '172.16.20.101', 'leader': 0}, {'status': 1, 'ip': '172.16.20.102', 'leader': 0}, {'status': 1, 'ip': '1
.16.20.110', 'leader': 0}, {'status': 1, 'ip': '172.16.20.103', 'leader': 0}]
```
(c)

*Figure 11: Faulty nodes in VLAN 20*

Fig a, b and c are part of the same output collected at leader $N_6$.

[see Fig. 11(a)] shows the result frame at N. This is the most recent and the last selected leader of the above depicted diagnosis cycle. As earlier discussed, [see Fig. 11(a)] also indicates the fault status and leader bits of all the nodes in VLAN 20. Please note that along with $N_6$, there are few more nodes which are marked as leaders. These nodes, when marked as leaders indicate that they were selected as leaders in the earlier stages of diagnosis cycle as they were found to be fault-free. This proves that DP-AFD ensures that every fault-free node gets a chance to become a leader at some point in the diagnosis cycle. This makes the fault diagnosis process a decentralized one. Thus, all the nodes together explore the network and are able to find the faulty nodes in the system.

In continuation of [see Fig. 11(a)], [see Fig 11(b)] depicts $N_6$ checking the leader status of every other node to ensure the whole network is explored. As shown in the

[see Fig 11(b)], every fault free node has become a leader whereas the faulty nodes have been detected. If in a situation where, let us say, N6 found a fault-free neighbour node $N_x$ with leader bit zero, then it sends its result frame to $N_x$. Then $N_x$ becomes the leader and continues the further diagnosis cycle. Whereas, if $N_x$ is not a neighbour of $N_6$, then $N_6$ ensures that the result frame is delivered to $N_x$ via intermediate leaders. $N_x$ also becomes the leader in the later part of the same diagnosis cycle. The last line of [see Fig. 11(b)] indicates all the ten nodes of VLAN 20 are checked by the most recent leader namely $N_6$. This ensures that all the nodes in the network are diagnosed.

[see Fig. 11(c)] shows $N_6$ indicating the fault status of every node in the VLAN 20. As $N_6$ is the most recent and the last leader of the network $N_6$ ensures the broadcasting of the fault diagnosis information.

**D. Scenario 4: Faulty nodes in VLAN 20 and VLAN 30**

The screenshot in [see Fig. 12] is captured at $N_9$ of VLAN 20. [see Fig. 12] shows transfer of diagnostic information between two VLANs namely VLAN 20 and VLAN 30. Here, the $N_9$ of VLAN 20 with IP address 172.16.20.109 has received the faulty nodes of VLAN 30. There are four faulty nodes in the network of VLAN 30 which are depicted in [see Fig. 12] along with their status bits as 1.

```
ip 172.16.20.109
received
received

FAUTLY NODES IN THE NETWORK From OTHER VLAN ARE {
{'status': 1, 'ip': '172.16.30.105', 'leader': 0}
{'status': 1, 'ip': '172.16.30.109', 'leader': 0}
{'status': 1, 'ip': '172.16.30.101', 'leader': 0}
{'status': 1, 'ip': '172.16.30.104', 'leader': 0}
```

*Figure 12: Inter VLAN faulty nodes*

The [see Fig. 13] shows the number of messages for different fault conditions in DP-AFD. As seen in the [see Fig. 13], the number of messages is reduced as the number of faults increase in the system. This is due to no response received by the leaders from the faulty nodes during the diagnosis cycle.

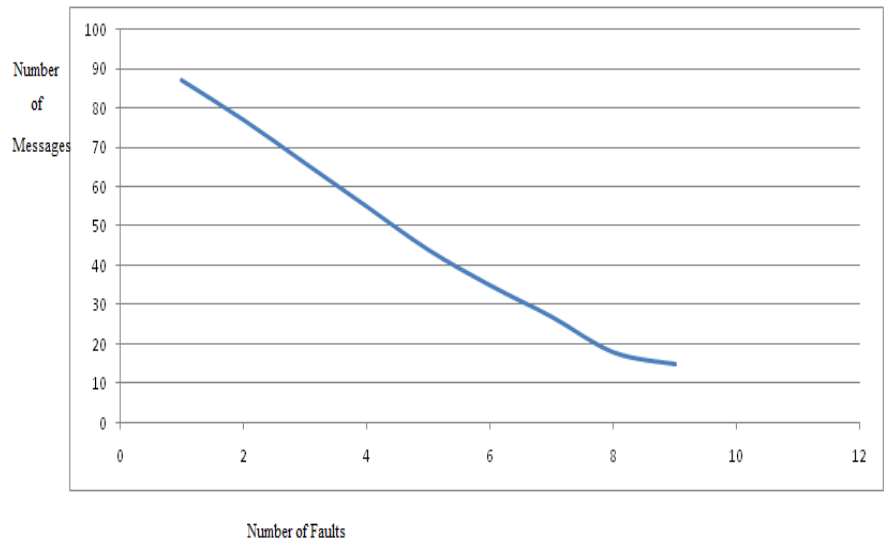

*Figure 13: Messages vs Faults*

## 6 Conclusion

DP-AFD algorithm is an innovative approach which attempts to reduce the number of communication messages thereby reducing the network bandwidth. Curbing the need of a central observer, our algorithm provides every node an equal chance to become a leader. This prevents system to be vulnerable to a single point of failure. It periodically checks the network and gives the accurate location of faults at the end of each diagnosis cycle. The new node and repaired node re-entry are allowed in the next diagnosis cycle. Each leader is elected on the basis of response time and follows a periodic approach. Such an algorithm can form a part of a network diagnostic tool which can be used for efficient network monitoring by the administrator. This algorithm can further be revised to make it event driven where new node entry and repaired node reentry can form events and they can trigger the algorithm execution.